\documentstyle[11pt]{article}
\begin{document}
\begin{titlepage}
\parindent 0.7cm
\centerline{\large\bf About the Mixing and CP Violation
in Neutrino System }

\vspace{1cm}
\centerline{ Yong Liu$^{1,2}$ and Utpal Sarkar$^3$ }
\vspace{0.5cm}
\centerline{1. Laboratory of Numeric Study for Heliospheric Physics }
\centerline{Chinese Academy of Sciences}
\centerline{P. O. Box 8701, Beijing 100080, P.R.China}
\centerline{2. Physics Department, National University of Singapore}
\centerline{Kent Ridge, Singapore 119260}
\centerline{E-mail: phyliuy@nus.edu.sg }
\centerline{3. Physical Research Laboratory, Ahmedabad 380 009, India }
\vspace{2cm}

\centerline{\large\bf Abstract}
\vspace{0.6cm}

Suppose the geometrical explaination to the weak CP phase in quark
sector is also valid for neutrinos, the mixing and CP violation in
neutrino system are discussed. We find a larger $J_{CP}$ than $3
\times 10^{-3}$ implys the large-mixing solution for solar neutrino
problem. In case of bi-maximal mixing, we predict relative large CP
violation with $J_{CP}$ larger than $10^{-3}$ in neutrino system,
except the third mixing angle approachs to $0$ or $\pi/2$ very
closely.
\\\\
PACS number(s): 11.30.E, 14.60.P, 12.15.F

\end{titlepage}

\centerline{\bf 1. Introduction}

In the Standard Model, photon and neutrino have no masses. However,
the masslessness of photon is ensured by gauge invariance while the
masslessness of neutrinos is only an artificial supposition. Since
it is found that, the observed solar neutrino fluxes are all below
the predictions based on the Standard Solar Model (SSM)
\cite{bahcall1}, and it is very difficult to explain these flux
deficits by modifying the SSM \cite{bahcall2}, people began to
guess that neutrinos may have non-zero masses, and can oscillate
from one flavor to another like that occuring in the sector of
quarks.

Recently, the Super-Kamiokande experiment has provided a strong
evidence for non-zero masses and oscillations of neutrinos
\cite{super1,super2}. Because it is the first sign for new physics
beyond the Standard Model, it has brought up a turbulent shock in
the research field of particle physics after the announcement of
the Super-Kamiokande result.

In analogy to the quark mixing in the Standard Model, it is expected 
that,
the mixing matrix of the neutrino sector has the similar structure to
that of quark sector. Then, it remind us to discuss the problem of 
mixing
and CP violation in neutrino system naturally
\cite{sarkar,schubert,gago,barger}.

Based on the postulation put forward by one of us and the collaborators
\cite{cgll,liu}, we investigate the mixing and CP violation in neutrino
system in this work. Here, we suppose that

A. The postulation on the relation between weak CP phase and the other
three mixing angles in Cabibbo-Kobayashi-Maskawa matrix for quark 
sector
be also available for neutrino system. Due to the similarity between 
quark
sector and lepton sector, we take this supposition as reasonable.

B. Neutrinos come in three families with no additional species, sterile
etc., and their masses are hierarchical with $m_{\nu_e}$ being smallest
and $m_{\nu_\tau}$ largest \cite{fishbane}.

C. Due to the confirmation of the Liquid Scintillator Neutrino Detector
(LSND) results at Los Alamos awaiting future experiments \cite{athana},
in the simplest explanation, solar neutrino data can be understood in
terms of $\nu_e-\nu_{\mu}$ oscillation with a small mass splitting not
to influence atmospheric data, and atmospheric data can be explained in
terms of $\nu_{\mu}-\nu_{\tau}$ large mixing with a large mass 
splitting
compared to the $\nu_e-\nu_{\mu}$ case \cite{barbieri}.

In fact, supposition A is necessary in this paper, while the other two
suppositions are only for the convenience. Although we need two mixing
angles precisely here, we need not limit which two of the three mixing
angles. The suppositions B and C are helpful to draw a clear physical
picture.

\vspace{0.5cm}
\centerline{\bf 2. The postulation on the weak CP phase}

In previous works \cite{cgll,liu}, we have postulated that, the weak CP
phase $\delta_{13}$ and the other three mixing angles $\theta_{12},
\theta_{23}$ and $\theta_{13}$ satisfy
\begin{equation}
\label{angle}
\sin\delta_{13}=\frac{ (1+s_{12}+s_{23}+s_{13})
                       \sqrt{1-s_{12}^2-s_{23}^2-s_{13}^2+
                       2 s_{12} s_{23} s_{13}} }{(1+
                       s_{12}) (1+s_{23}) (1+s_{13})}
\end{equation}
where, the convention $s_{ij}=\sin\theta_{ij}, c_{ij}=\cos\theta_{ij}$
(the "generation" labels $i,j=1,2,3$) are used and, $\delta_{13}$ and
$\theta_{ij}$ are those present in the standard parametrization of
the Cabibbo-Kaboyashi-Maskawa (CKM) matrix
\begin{equation}
V_{KM}= \left (
\begin{array}{ccc}
   c_{12} c_{13} & s_{12} c_{13}& s_{13} e^{-i \delta_{13}} \\
   -s_{12} c_{23}-c_{12} s_{23} s_{13} e^{i \delta_{13}} &
   c_{12} c_{23}-s_{12} s_{23} s_{13} e^{i \delta_{13}}    &
   s_{23} c_{13}\\
   s_{12} s_{23}-c_{12} c_{23} s_{13} e^{i \delta_{13}}  &
   -c_{12} s_{23}-s_{12} c_{23} s_{13} e^{i \delta_{13}} &
   c_{23} c_{13}
\end{array}
\right )
\end{equation}
with the real
angles $\theta_{12}, \theta_{23}$ and $\theta_{13}$ can all be made to
lie in the first quadrant. The phase $\delta_{13}$ lies in the range
$0<\delta_{13}<2 \pi$. In following, we will fix the three angles
$\theta_{12}, \theta_{23}$ and $\theta_{13}$ in the first quadrant.

The geometry meaning of Eq.(\ref{angle}) is evident.
$\delta_{13}$ is the solid
angle enclosed by $(\pi/2 -\theta_{12}), (\pi/2-\theta_{23})$ and
$(\pi/2-\theta_{13})$ standing
on a same point, or, the area to which the solid angle
corresponding on a unit spherical surface.

Hence, to make $(\pi/2 -\theta_{12}), (\pi/2-\theta_{23})$ and
$(\pi/2-\theta_{13})$ be able to enclose a solid angle, the following
relation must be hold.
\begin{equation}
\label{tri}
(\frac{\pi}{2}-\theta_{ij})+(
\frac{\pi}{2}-\theta_{jk}) \geq (
\frac{\pi}{2}-\theta_{ki})  \;\;\;\;\;\;
(i\not=j\not=k\not=i=1,2,3. \;\;\;\theta_{ij}=\theta_{ji})
\end{equation}

Eq.(\ref{tri}) and Eq.(\ref{angle}) are the most important constraints
in this work, on which the following discussions are based.

\vspace{0.5cm}
\centerline{\bf 3. The relevant experimental results
on neutrino masses and mixing}

The recent analysis made by Hata and Langacker \cite{hata2} gives 
viable
solutions for the BAHCALL SSM \cite{bahcall3,PDG}. With the
Mikheyev-Smirnov-Wolfenstein (MSW) mechanism being
considered \cite{wolf,mikh}, they give the small-mixing solution
\begin{equation}
\label{emu1}
\delta m_{sol}^2 \sim 5\times 10^{-6} eV^2 \;\;\;\;\;\;
sin^2 2\theta_{sol} \sim 8\times 10^{-3}
\end{equation}
and the large-mixing solution
\begin{equation}
\label{emu2}
\delta m_{sol}^2 \sim 1.6\times 10^{-5} eV^2 \;\;\;\;\;\;
sin^2 2\theta_{sol} \sim 0.6.
\end{equation}
Vacuum oscillation also provide solutions
\begin{equation}
\label{emu3}
\delta m_{sol}^2 = (5\sim 8)\times 10^{-11} eV^2 \;\;\;\;\;\;
sin^2 2\theta_{sol} = 0.65 \sim 1.
\end{equation}

The atmospheric neutrino data from Super-Kamiokande etc. imply that the
parameters of the $\nu_{\mu}-\nu_{\tau}$ oscillation of the atmospheric
neutrinos are \cite{super1,super2,PDG}
\begin{equation}
\label{mutau}
10^{-4} \leq \delta m_{atm}^2 \leq 10^{-2} eV^2 \;\;\;\;\;\;
sin^2 2\theta_{atm} \approx 1.
\end{equation}

We have known that, the small-mixing solution causes the 
energy-spectrum
distortion while the large-mixing solution causes the day-night flux
difference, and, the vacuum-oscillations cause seasonal variation of
the $^7B_e$ solar neutrino flux \cite{PDG}.

In the next section, we will see that, with the Super-Kamiokande
results about the $\nu_{\mu}-\nu_{\tau}$ oscillation being
admitted, the small-mixing solution differents from the
large-mixing solution in CP violation greatly. Because the
discussions here are only related to the mixing angles, and the MSW
large-mixing solution gives about the same mixing as that given by
the vacuum oscillation solution, so, we only discuss the two cases
of small- and large-mixing as indicated by the MSW solutions.

\vspace{0.5cm}
\centerline{\bf 4. Some predictions on the mixing and CP violation
in neutrino system}

Let us return to the constraint Eq.(\ref{tri}) and recall the clear
geometry meaning of Eq.(\ref{angle}), what can we extract from them?
As we have supposed in section 1, Eq.(\ref{mutau}) tells us
\begin{equation}
\label{pi4}
\theta_{\mu \tau}\approx \pi/4.
\end{equation}
Keep this point in mind, then, by the use of Eq.(\ref{tri}),
Eq.(\ref{emu1}-\ref{emu3}) will give restriction on the mixing
angle between $\nu_e$ and $\nu_\tau$. In the meantime, Eq.(\ref{angle})
tell us some information on the CP violation in neutrino system. Let
us talk more detail about these two problems.

(1) About the mixing angle between $\nu_e$ and $\nu_\tau$.

From Eq.(\ref{tri}), we have
\begin{equation}
\label{atr}
|(\frac{\pi}{2}-\theta_{e \mu})-(
\frac{\pi}{2}-\theta_{\mu \tau})| \leq (
\frac{\pi}{2}-\theta_{e \tau}) \leq
Min ( \pi/2, \;\; (\frac{\pi}{2}-\theta_{e \mu})+(
\frac{\pi}{2}-\theta_{\mu \tau}) )
\end{equation}
Note that, Eqs.(\ref{emu1}-\ref{emu2}) imply
for small- and large-mixing solutions
$$
\theta_{e \mu}\sim 0.045 \;\;\; {\rm or} \;\;\; \pi/2-0.045
$$
and
$$
\theta_{e \mu}\sim 0.443 \;\;\; {\rm or} \;\;\; \pi/2-0.443
$$
respectively.
Considering Eq.(\ref{pi4}), then we obtain
\begin{equation}
\label{etau1}
0 \leq \theta_{e \tau} \leq \pi/4+0.045
\end{equation}
or
\begin{equation}
\pi/4-0.045 \leq \theta_{e \tau} \leq  \pi/4+0.045
\end{equation}
for the case of small-mixing solution. And
\begin{equation}
0 \leq \theta_{e \tau} \leq \pi/4+0.443
\end{equation}
or
\begin{equation}
\label{etau2}
\pi/4-0.443 \leq \theta_{e \tau} \leq \pi/4+0.443
\end{equation}
for the case of large-mixing solution.
Eq.(\ref{etau1}-\ref{etau2}) are the constraints on the mixing angle
$\theta_{e \tau}$.

(2) About the CP violation in neutrino system.

Now, we discuss the CP violation in neutrino system. As is well known,
all the CP violated observables are proportional to the Jarlskog 
invariant
\cite{jarlskog,paschos}
\begin{equation}
\label{jarl}
J_{CP}=s_{12} s_{13} s_{23} c_{12} c_{13}^2 c_{23} s_{\delta_{13}}
\end{equation}
where, $s_{\delta_{13}}\equiv sin \delta_{13}$ with $\delta_{13}$
the weak CP phase presenting in the CKM matrix of the lepton
sector. And, we instead $\theta_{12}, \theta_{13}$ and
$\theta_{23}$ by $\theta_{e \mu},
\theta_{e \tau}$ and $\theta_{\mu \tau}$ respectively in following.

With $\theta_{\mu \tau}$ given by the Super-Kamiokande data
definately, the small- and large-mixing solutions and the vacuum
oscillation will give the pemitted ranges for $\theta_{e \tau}$
correspondingly. Substitute Eq.(\ref{angle}) into Eq.(\ref{jarl}),
we obtain $J_{CP}$ as a function of $\theta_{e \tau}$. Then, we can
draw the curve which $J_{CP}$ versus $\theta_{e \tau}$. The results
are shown in Fig.(1). From the figure, we find that, the
small-mixing solution corresponds to small CP violation while the
large-mixing solution corresponds to large CP violation.

For the small-mixing solution, $J_{CP}$ is very small. The maximum of
$J_{CP}$ is about $5\times 10^{-3}$ when $\theta_{e \mu}$ nearing to
$0$, and $1.5\times 10^{-4}$ when $\theta_{e \mu}$ nearing to $\pi/2$.

For the large-mixing solution, $J_{CP}$ is relative large. The
maximum of $J_{CP}$ is about $1.5\times 10^{-2}$ when $\theta_{e
\mu}$ nearing to $0.443$, and $3.2\times 10^{-2}$ when $\theta_{e
\mu}$ nearing to $\pi/2-0.443$.

Now, it is evident that, if the future experiments on the CP
violation in neutrino system tell us that $J_{CP}$ is larger than
$5\times 10^{-3}$, the mixing between $\nu_e$ and $\nu_\mu$ must be
large. Recalling the same observable in quark sector obtained via
$K^0-\overline{K^0}$ system is about $10^{-4}$, if it is the same
order of magnitude in the neutrino system, then, the mixing between
$\nu_e$ and $\nu_\tau$ is either around $0.8$ with a relative
narrow window or nearing to $0$ or $(\pi/4+0.443)$ very closely.

On the other hand, either $\theta_{e \mu}$ takes $\sim 0.443$ or it
takes $\sim (\pi/2-0.443)$ for large-mixing solution can also be
distingushed to some extent. For example, if experiment tells us
$J_{CP}>0.015$, it must be $\theta_{e \mu} \sim 0.443$.

Finally, with the CHOOZ result \cite{chooz} being considered, that
is, ${\rm sin}^2 2 \theta_{e \tau}< 0.2$, then we have
$$
0<\theta_{e \tau}<0.23 \;\;\;\; {\rm or} \;\;\;\;
\pi/2-0.23<\theta_{e \tau}<\pi/2.
$$
Based on the above constraint, we can see from Fig.(1), firstly,
the small-mixing solution with $\theta_{e \mu} \sim (\pi/2-0.045)$
and the large-mixing solution with $\theta_{e
\mu} \sim (\pi/2-0.443)$ have been excluded. Secondly, the possible
domain $\pi/2-0.23<\theta_{e \tau}<\pi/2$ should also be
eliminated. Thirdly, $J_{CP}$ can still be large to about $0.02$ in
the case of large-mixing solution with $\theta_{e \mu} \sim 0.443$,
especially, the larger $J_{CP}$ than $3 \times 10^{-3}$ will
exclude the possibility of small-mixing solution finally.

Maybe, the most interesting conclusion is about the bi-mixmal mixing
\cite{bpww,tanimoto}.
From the above analysis, we can see that, in most of the permitted 
range
of the third angle - $\theta_{e \tau}$, there will be a relative
large CP violation.

Suppose that $\theta_{e \mu}=\theta_{\mu \tau}=\pi/4$, similarly,
we can draw the curve which $J_{CP}$ versus $\theta_{e \tau}$ in
the permitted range of $\theta_{e \tau}$. The result is shown in
Fig.(2). We find that, except for $\theta_{e \tau}$ nearing to $0$
or $\pi/2$, $J_{CP}$ is larger than $10^{-3}$ in most of the
permitted range of $\theta_{e \tau}$. And, the maximum of $J_{CP}$
can reach to about $0.018$ when the CHOOZ result is considered.

\vspace{0.5cm}
\centerline{\bf 5. Conclusions}

Starting from the postulation on the relation between weak CP phase
and the other three mixing angles in the CKM matrix, we have 
investigated
the mixing and CP violation in the neutrino system.

We suppose that, the solar neutrino problem be understood in terms
of $\nu_e-\nu_{\mu}$ oscillation with a small mass splitting. With
the definite large mixing between $\nu_{\mu}-\nu_{\tau}$ indicated
by the Super-Kamiokande data, and the CHOOZ result being
considered, we obtain the relevant constraints on the mixing
between $\nu_e$ and $\nu_\tau$. We find, $0 \leq \theta_{e
\tau} \leq 0.23$ is permitted by the small- and the large-mixing
solutions.

Besides, the mixing between $\nu_e$ and $\nu_\mu$ is limited as
$\theta_{e \mu} \sim 0.045$ for the small-mixing solution or
$\theta_{e \mu} \sim 0.443$ for the large-mixing solution. And, a
larger $J_{CP}$ than $3 \times 10^{-3}$ will finally exclude the
possibility of small-mixing solution.

Furthermore, if the suppositions B and C in section {\bf 1} holds,
a $J_{CP}$ larger than $3 \times 10^{-3}$ implys the large-mixing
solution for solar neutrino problem. And, if it takes the same
order for $J_{CP}$ in the neutrino system as the one in quark
system, the mixing between $\nu_e$ and $\nu_\mu$ will be very small
$(\sim 10^{-2} {\rm \; or \;less})$.

For the case of bi-maximal mixing, we predict a large
CP violation in neutrino system with $J_{CP}$ larger than $10^{-3}$,
except the third mixing angle approachs to $0$ or $\pi/2$ very closely.

Finally, although we have made some suppositions in this work, the
basis and the method used here is actually valid for a more general
discussion.\\\\
{\bf Notes}: In fact, this work has been finished and submitted
before the last July. Due to some reason, we have not put it on the net
in time. Just two days ago, when we noted the paper hep-ph/0004020 by
Sin Kyu Kang, C. S. Kim and J. D. Kim and found their results are almost
the same as those of us, although they based on some concrete model while
we only started out from our postulation, we are encouraged to post this
short paper.

\newpage

\begin{figure}[htb]
\mbox{}
\vskip 7in\relax\noindent\hskip -1 in\relax
\includegraphics{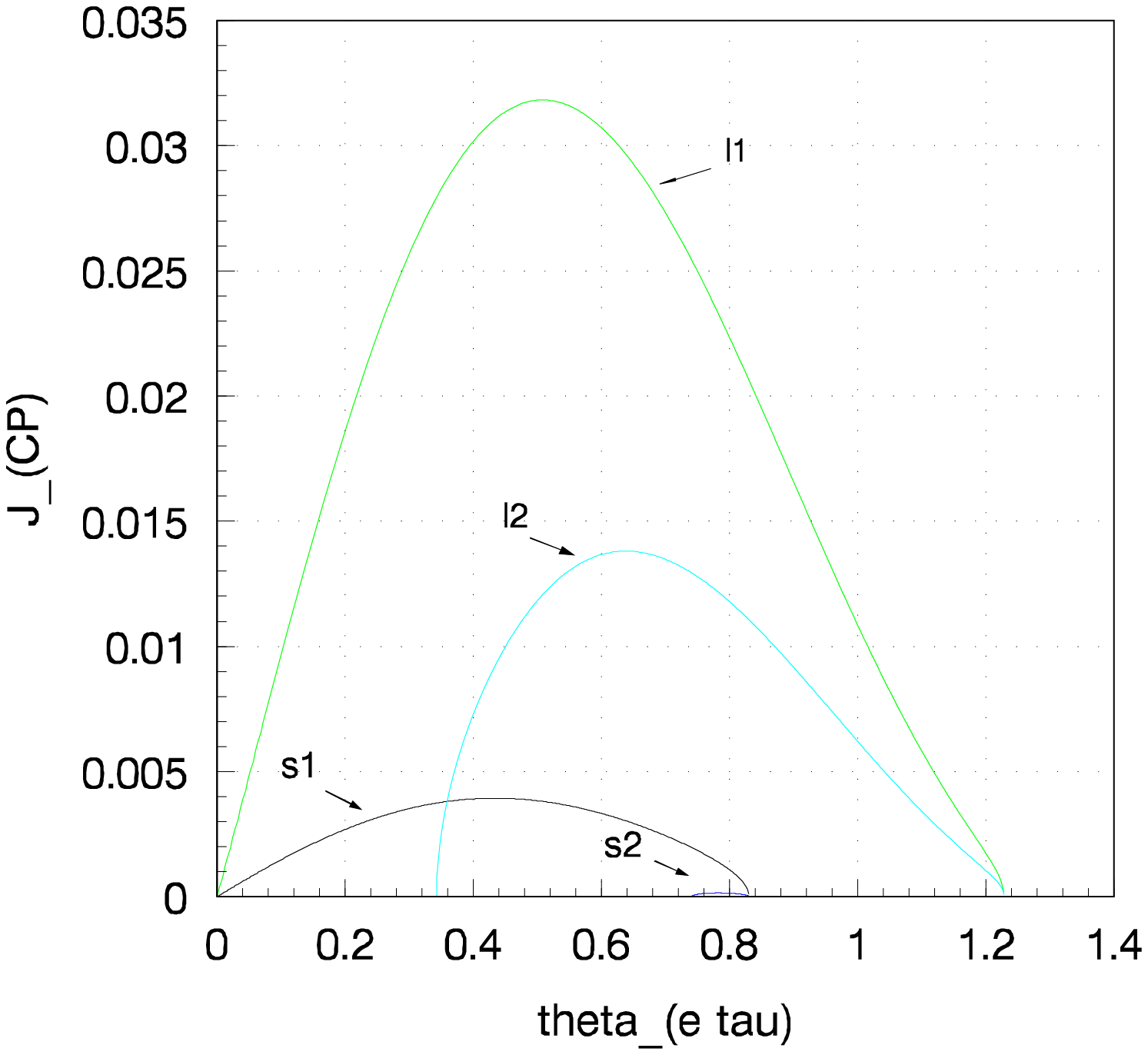}
\caption{
$J_{CP}$ versus $\theta_{e \tau}$. Where,
$\theta_{\mu \tau}=\pi/4$. The curves s1, s2, l1 and l2 corresponds to
the cases of
$\theta_{e \mu}=0.045,\; (\pi/2-0.045),
\; 0.443$ and $(\pi/2-0.443)$ respectively.
}
\end{figure}

\begin{figure}[htb]
\mbox{}
\vskip 7in\relax\noindent\hskip -1 in\relax
\includegraphics{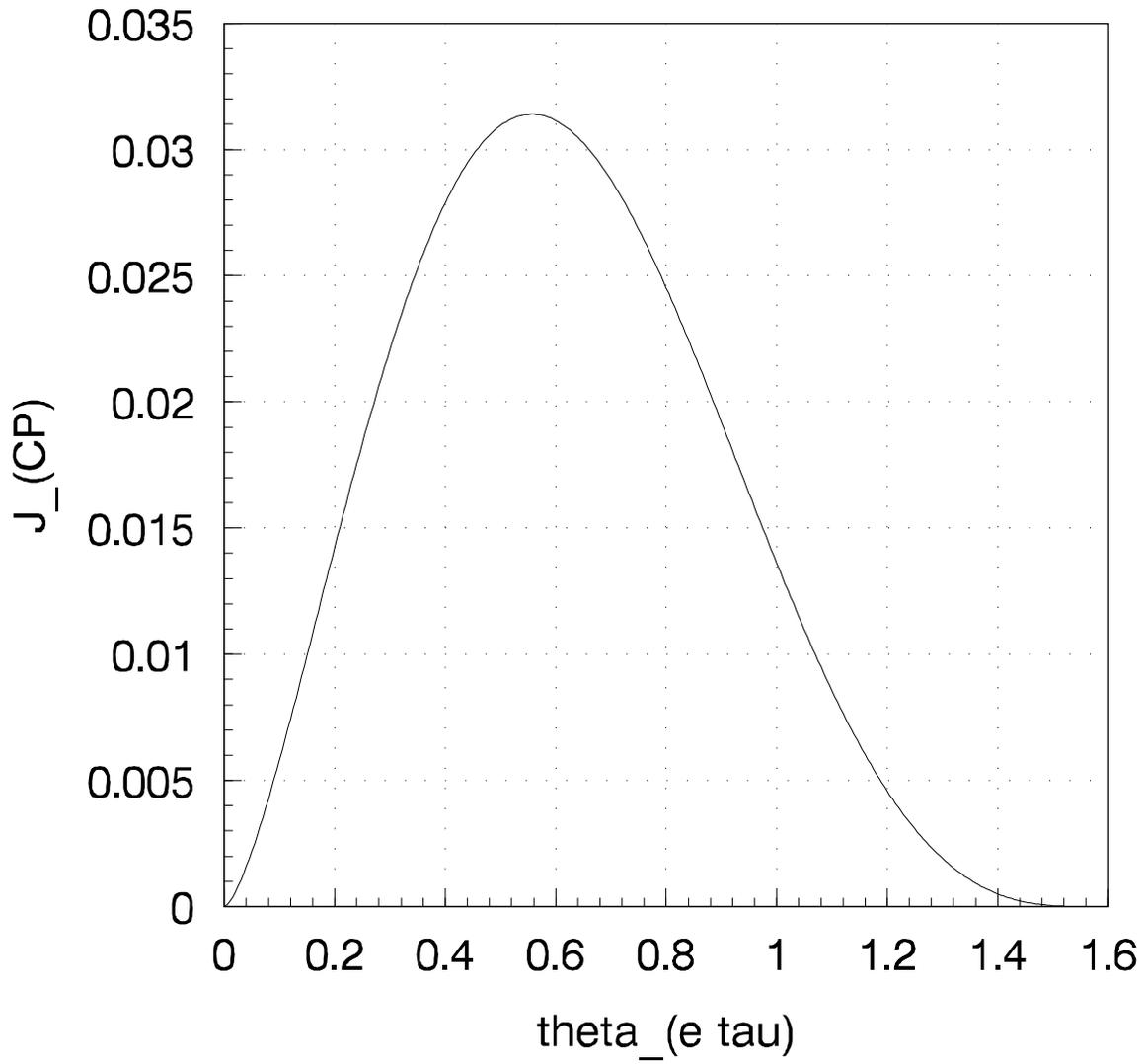}
\caption{
$J_{CP}$ versus $\theta_{e \tau}$. Where,
$\theta_{e \mu}=\theta_{\mu \tau}=\pi/4$. So, the
permitted range for $\theta_{e \tau}$ is $0\sim \pi/2$.
}
\end{figure}

\end{document}